\documentclass[12pt]{article}

\begin{document}
\begin{center}
\Large Ordering the universe with naked-eye observations

\vspace{1.0cm}

\normalsize
{\it Alan B. Whiting \\
University of Birmingham}
\end{center}
\vspace{1.0cm}
\large
The geocentric universe, in its most developed form as set out by Ptolemy,
was a remarkably successful and coherent theory.  It did not, however,
specify the order of the planets, that is, which was closer to Earth
and which farther away.  One would naively think that seeing one planet
pass in front of another would settle the matter.  In practice such
mutual phenomena happen too rarely for them to have been useful.
Even in principle, it turns out that most naked-eye observations of a central
event would show nothing conclusive, with the exception of some occultations
by Venus that would demonstrate it to be the lowest (nearest) planet.
However, if one's theory were good enough to allow conclusions from
{\it not} seeing changes, one could find that Mars is probably lower
than Jupiter and Saturn, and possibly that the overall order is
Venus-Mars-Jupiter-Saturn-Mercury.
\normalsize

\vspace{1.0cm}

\noindent{\it The order of the universe}

The geocentric theory of the universe, as most highly developed by
the Greco-Roman astronomer
 Ptolemy, was a great achievement of mathematical astronomy
and held the field for several
centuries.  However, one thing it did not specify was the ordering
of the planets.  Clearly, the Moon was closest to Earth (lowest)
because it occulted all other planets (as well as stars); but
otherwise the theory was ambiguous.  The
positions of each of the moving bodies could be calculated without
reference to the others.  There was a consensus that the
slower-moving planets should be farther away, but even then the
relative heights of Mercury, Venus and the Sun were unknown,
since their average motions were the same.

Prompted by the 2021 Great Conjunction of Jupiter and Saturn, 
I wondered whether an occultation of the latter by the former
(instead of just a close passage) could have given some
information\footnote{The term `occulation' is used when the
body passing in front is larger than the one behind, `transit'
when the reverse is true.  Since the planets are of roughly
similar sizes, either might be used for planet-planet
mutual phenomena; here I stay with `occultation.'}.
Obviously, a telescope would show one passing in front of the other,
but such an instrument was not available to the ancients.  Instead,
a naked-eye observer would be limited to observing changes in
brightness.

However, a calculation of actual events shows that they happen far too
rarely to have been useful to ancient astronomers.  In a compilation
covering the years 1557 to 2230$^1$, S. C. Albers found three possibly
observable Venus-Jupiter events (rejecting those too close to the Sun);
two observable events involving Mercury; and two observable Mars-Jupiter
events.  For astronomers in a limited region of Earth, these seven events
would be reduced to at most two, possibly none.  There was a complete lack
of Jupiter-Saturn events.  Even without actually performing the calculations
for the period whose observations were available to Ptolemy (roughly
750 BCE to 150 CE), we can be confident that at best a handful of
events could have been seen.  It is not suprising that none are mentioned
in the \emph{Almagest}$^2$.

Let us change the question.  Supposing an unlimited time of observation,
or some reliable record of mutual events handed down over time, what could
an ancient geocentric astronomer {\it in principle} have concluded from
mutual planetary events?

\vspace{0.5cm}
\noindent{\it Data}

For the following calculations I used mean distances, speeds, sizes and
magnitudes of planets from a textbook at hand$^3$ plus data from a
reasonably recent almanac$^4$.  Since the object is a set of indicative
calculations of generic phenomena, I sought no great accuracy; 
complications like limb darkening and the aspect of Saturn's rings are
ignored.  For transits involving only outer planets, calculations were
done at opposition.  For those involving inner planets, calculations
were done at Mercury's mean greatest elongation, $22.8^{\rm o}$, taken
as about the effective limit of observability.  
Input data are listed
in Table \ref{data}.

\begin{table}
\begin{tabular}{lcccc}
& \multicolumn{2}{c}{$22.8^{\rm o}$ or near}& \multicolumn{2}{c}{Opposition or far} \\
Planet & Magnitude & Size (arcsec) & Magnitude & Size (arcsec) \\ \hline
Mercury & -1.7 & 7.3 & & \\
Venus & -0.1 & 53.6 & -3.6 & 10.9 \\
Mars & 1.9 & 3.9 & -2.01 & 17.9 \\
Jupiter & -1.5 & 38.0 & -2.70 & 46.9 \\
Saturn & 0.6 & 15.9 & 0.2 & 19.4 \\
\end{tabular}
\caption{Basic data for planet mutual phenomena.  
Mercury is only listed at its greatest elongation;
Venus is listed at
near and far elongations of $22.8^{\rm o}$, the outer planets at
that elongation and opposition.}
\label{data}
\end{table}

\vspace{0.5cm}
\noindent{\it Calculations}

All transits were assumed to be central.  
A naked-eye observer would see two planets approach each other
until they formed a single object, with a combined brightness
given in the calculations.  As one covered the other it would
dim over the period shown, remain at that lower level for
the duration of `totality,' then recover.

What can a visual observer detect?  There is anecdotal evidence
of exceptional variable-star observers who can make estimates
good to 0.05 magnitude.  However, they are rare, and of course
have calibrated comparison stars to use.  As a rule of thumb
we can probably take a dip of 0.1 magnitude to be the limit
of what an experienced observer might notice, with the caveat
that if it happens slowly, or with nothing of similar brightness
to serve as a reference, the threshold for detection could be
much larger.

The results for outer planet events are given in Table \ref{outer}.
The case of Jupiter occulting Saturn, the original prompt for this
study, has a dip of 0.07 magnitude happening over hours.  It would
surely pass unnoticed.  Indeed, variations in brightness due to
changing airmass alone would mask it.  Phenomena involving Mars
are marginally better, but would still show nothing to the naked-eye
observer.

\begin{table}
\begin{tabular}{ccccc}
Planets & Magnitude & Magnitude & Ingress & Totality \\
 & outside & totality & minutes & minutes \\
Mars on & -3.16 & -3.05 & 32 & 52 \\
Jupiter & & & & \\
Mars on & -2.14 & -2.03 & 26 & 2 \\
Saturn & & & & \\
Jupiter on & -2.77 & -2.70 & 143 & 203 \\
Saturn & & & & \\
\end{tabular}
\caption{Outer planet mutual phenomena}
\label{outer}
\end{table}

For the inner planets there are two possibilities with
Mercury at greatest elongation: Venus on
the near side of its orbit could occult Mercury, and Mercury
could occult Venus on the far side of the latter's orbit.
Results are shown in Table \ref{inner}.  For a near Venus,
the event would be very clear: a large, mostly unlit crescent Venus
covering Mercury gives an unmistakable drop in light, and
does it rather quickly.  (Indeed, if we could arrange this
to happen near Sirius, at mag. -1.6, it could be
spectacular.)  Venus, then, is clearly closer than
Mercury.  The other possibility is more equivocal.  A drop
to about half the original brightness in a few minutes
would seem to be clearly observable.  However, there are no
stars nearly this bright to use as comparisons.  It would
certainly be different from the other situation
observationally.

\begin{table}
\begin{tabular}{ccccc}
Planets & Magnitude & Magnitude & Ingress & Totality \\
&outside & totality & minutes & minutes \\
Venus (near) & -1.92 & -0.7 & 3.8 & 24 \\
on Mercury & & & & \\
Mercury & -3.77 & -3.25 & 12 & 5.8 \\
on Venus (far) & & & & \\
\end{tabular}
\caption{Inner planets mutual phenomena}
\label{inner}
\end{table}

The final group of mutual phenomena occur when the inner planets pass in
front of outer planets, with results shown in Table 4.
The near-side occultations by Venus of Jupiter would show that
Venus is the lower planet pretty clearly; with Saturn, it is less
evident.  For Mars, the drop in brightness
would be on the edge of perception, though perhaps if it happened near
a suitable comparison star it might be seen.  Occultations by Mercury of
the outer planets show nothing perceptible, as are the situations with
Venus on the far side of
its orbit passing in front of them.

\vspace{0.5cm}

\begin{table}
\begin{tabular}{ccccc}
Planets & Magnitude & Magnitude & Ingress & Totality \\
&outside & totality & minutes & minutes \\
Venus (near) & -0.26 & -0.1 & 1.7 & 22 \\
on Mars & & & & \\
Venus (near) & -1.76 & -0.1 & 34 & 14 \\
on Jupiter & & & & \\
Venus (near) & -0.56 & -0.1 & 19 & 44 \\
on Saturn & & & & \\
Mercury & -1.74 & -1.7 & 5.5 & 4.8 \\
on Mars & & & & \\
Mercury & -2.36 & -2.34 & 3.8 & 16 \\
on Jupiter & & & & \\
Mercury & -1.82 & -1.80 & 3.4 & 3.9 \\
on Saturn & & & & \\
Venus (far) & -3.61 & -3.60 & 3.0 & 5.3 \\
on Mars & & & & \\
Venus (far) & -3.75 & -3.73 & 4.3 & 11 \\
on Jupiter & & & & \\
Venus (far) & -3.62 & -3.61 & 3.9 & 1.8 \\
on Saturn & & & & \\
\end{tabular}
\label{mixed}
\caption{Mutual phenomena, inner planets on outer}
\end{table}

\vspace{0.5cm}

\noindent{\it Analysis}

Far from being a sure way to order the geocentric universe, as one might
naively think, planets passing in front of each other turn out to be
generally uninformative to the naked eye.
The only clear observations would show Venus to
be lower than Mercury and Jupiter, and possibly Saturn and Mars. 

There is a possibility of learning more, though it requires us to take
a further (and much larger) step away from actual history.  Consider the
case of Mars transiting Saturn.  The combined object would retain its
red colour throughout the event.  One might reason, then, that Mars
could not have been covered, and thus was nearer.

This is predicated on the assumption that the nearer planet did in fact
pass in front of the more distant one.  For this one must have predictions
accurate to within a planetary diameter, a matter of seconds of arc.
Now, the accuracy of the Ptolemaic theory varies with the position of the
planet in its orbit and with the observations used to calibrate it.
An indication of the uncertainty of the actual theory is
given by two instances in which a planet is declared to have
occulted a star (Ref. 2, p. 477 n. 17 and p. 522 n. 16) when modern
calculations show a miss by 12 and 15 arc minutes.  For predicting
transits, then, it was not up to the job.  Ancient observers would be
in a position analogous to that of W. H. Smyth examining 
the close double star $\zeta$ Herculis.  He reports$^5$ that
during one apparition he saw a red spot on the disc
of the primary, which he took to be the fainter and redder secondary
star in transit. But lacking any orbital prediction or feel for how
large the disc of a star actually was, he could not be sure, and
 alternatively says it could have been `a spurious image or colour.' 

However, if we postulate a theory (or some other condition) that allows us
to be confident that an occulation did occur and thus
to learn from not seeing anything happen, more conclusions are possible.
As noted, we could conclude that Mars is lower than Jupiter or Saturn by
the colour of its transits. 
And if, say, our naked-eye observer sees Jupiter shine as
brightly as ever as it reaches Saturn, he might conclude that Jupiter must
be the nearer.  Similarly, the failure of Mercury to dim any of the outer
planets could be interpreted as it lying at a higher altitude, a distinctly
unsettling idea for the fast-is-close consensus.
With these additions, our geocentric universe
has the planets in the order Venus-Mars-Jupiter-Saturn-Mercury.

\vspace{0.5cm}

\noindent{\it Conclusions}

Seeing one planet pass in front of another turns out to be generally
uninformative to the naked eye.  Historically, it happens too rarely
to be of interest.  Even in principle, it could only show that Venus is
the closest planet beyond the Moon.  In retrospect, this is not surprising.
The nearer planets are generally closer to the Sun and thus of higher
surface brightness, so the occulted planet contributes less to their
combined light.  The exception, Mercury, has little effect because it's
so small.  

I am indebted to an anonymous referee for the suggestion of a set of
phenomena that provides an illuminating contrast:
mutual events of Jupiter's Galilean satellites.  Of course these 
are
not visible to the naked eye, but in binoculars and small telescopes
the moons show no discs and observations are similarly limited to
changes in brightness.  Such observations are more useful than
planet mutual events for two main reasons.  First, they are far
more common, hundreds happening every six years as Earth passes
through their orbital plane.  Second, the moons are of similar
surface brightness, so the combined light typically
drops by a half-magnitude or more, easily noticed by comparison
with other moons.  In addition, there are eclipses, which are impossible
for planet-planet mutual phenomena.  The Observatoire de Paris coordinates
campaigns to use visual observations to update the moons'
ephemerides$^6$.

For planet mutual phenomena, it's only if we postulate an
anachronistically accurate method of prediction and an unhistorical
run of events that we could put the geocentric planets in order.
It is ironic that the invention of the telescope,
which alone could have made mutual phenomena useful for the geocentric universe,
instead proved its undoing.
	
\vspace{1.0cm}
\begin{center}
{\it References}
\end{center}

\noindent (1) Steven C. Albers, \emph{Mutual Occultations of Planets:
1157 to 2230}, Sky and Telescope \textbf{57, 3}, 220, 1979 \\
\noindent (2) G. J. Toomer, \emph{Ptolemy's Almagest}
(Princeton University Press, Princeton), 1998 \\
\noindent (3) H. Karttunen, \emph{et al.}, (eds),
\emph{Fundamental Astronomy, 3rd ed.} (Springer, London), 1996 \\
\noindent (4) U. S. Naval Observatory, \emph{Nautical Almanac for the
year 2016}, (U. S. Government Printing Office, Washington), 2015 \\
\noindent (5) William H. Smyth, \emph{The Bedford Catalog}
(Willmann-Bell reprint, Richmond), 1986, p. 370 \\	
\noindent (6) www.imcce.fr/recherche/campagnes-observations/phemus/phemu
\end{document}